\documentclass[runningheads,twocolumn]{llncs}

\usepackage{graphicx}
\usepackage{float}
\usepackage{amsmath}
\usepackage{subcaption}
\usepackage{xurl}
\usepackage{array}

\usepackage{xcolor}

\usepackage{etoolbox}
\usepackage{setspace}
\usepackage[margin=1in]{geometry}

\usepackage{fancyhdr}
\AtBeginEnvironment{tabular}{\small}

\DeclareCaptionFormat{custom}
{%
    \textbf{#1#2}\textit{\small #3}
}
\usepackage{bold-extra}

\usepackage{caption}
\begin{document}

\twocolumn[
\begin{@twocolumnfalse}

\noindent\rule{\textwidth}{0.4pt}
\vspace{-6mm}

\title{{\Huge \bf \scshape Leveraging Machine Learning for Multichain DeFi Fraud Detection\endgraf\rule{\textwidth}{.4pt}}}

\author{\large \normalfont Georgios Palaiokrassas\inst{*,1,2} \and
Sandro Scherrers\inst{2,3} \and 
Iason Ofeidis\inst{1,2} \and 
Leandros Tassiulas\inst{1,2}}

\authorrunning{G. Palaiokrassas, S. Scherrers, I. Ofeidis  and L. Tassiulas}

\institute{\small \normalfont \itshape Department of Electrical Engineering, Yale University, USA \and
Yale Institute for Network Science, USA \and
School of Management, Yale University, USA 
}

\maketitle

\begin{abstract} 
Since the inception of permissionless blockchains with Bitcoin in 2008, it became apparent that their most well-suited use case is related to making the financial system and its advantages available to everyone seamlessly without depending on any trusted intermediaries. Smart contracts across chains provide an ecosystem of decentralized finance (DeFi), where users can interact with lending pools, Automated Market Maker (AMM) exchanges, stablecoins, derivatives, etc. with a  cumulative locked value which had exceeded 160B USD. While DeFi comes with high rewards, it also carries plenty of risks. Many financial crimes have occurred over the years making the early detection of malicious activity an issue of high priority. The proposed framework introduces an effective method for extracting a set of features from different chains, including the largest one, Ethereum and it is evaluated over an extensive dataset we gathered with the transactions of the most widely used DeFi protocols (23 in total, including Aave, Compound, Curve, Lido, and Yearn) based on a novel dataset in collaboration with Covalent. Different Machine Learning methods were employed, such as XGBoost and a Neural Network for identifying fraud accounts detection interacting with DeFi and we demonstrate that the introduction of novel DeFi-related features, significantly improves the evaluation results, where Accuracy, Precision, Recall, F1-score and F2-score where utilized.  
\end{abstract}

{\scriptsize } 
\vspace{.1cm}

\noindent\rule{\textwidth}{0.4pt}

\vspace{.5cm}

\end{@twocolumnfalse}
]

\section{Introduction}

Blockchain was originally proposed in 2008 by Nakamoto \cite{nakamoto2008bitcoin}, as the accounting method for Bitcoin cryptocurrency.
In the years that followed the technology and ideas behind blockchain evolved, more blockchains were introduced, and its potential for different applications beyond cryptocurrency became apparent such as for IoT applications \cite{palaiokrassas2021combining}, identity management \cite{liu2020blockchain}, copyrights management \cite{palaiokrassas2019deploying}, decentralized finance \cite{bartoletti2021sok}, machine learning \cite{li2022survey} and more.
A milestone for the course of blockchain technology was the development of the Ethereum project, offering new solutions by enabling smart contracts’ implementation and execution \cite{wood2014ethereum}. 

Ethereum smart contracts are actually program instances running on the decentralized network and allow further the construction of sophisticated on-chain financial systems, namely Decentralized Finance (DeFi). 
In the DeFi ecosystem, any entity can deploy a financial protocol, by implementing the respective smart contracts and deploying it on the Ethereum blockchain network \cite{qin_empirical_2021}.
Users can interact with lending pools such as Aave and Maker
, Automated Market Maker exchanges, stablecoins, derivatives, and asset management platforms. Smart contracts are now also available on a variety of other blockchains, which we have included in this research.

Several works have combined machine learning and blockchain focusing on both Bitcoin and Ethereum platforms, addressing problems such as Ponzi Scheme detection \cite{zhang2021detecting}, address clustering \cite{wu2022tutela}, \cite{beres2021blockchain}, \cite{victor2020address}, cryptocurrency price prediction \cite{jang2017empirical}, phishing detection \cite{kabla_eth-psd_2022}, illicit accounts detection \cite{li2022survey} and many more. 
Existing works on Fraud Detection in Bitcoin and Ethereum mainly focus on proposing Machine Learning methods, utilizing labeled datasets for evaluation, and extracting features from the transaction history. In this direction, we extend existing approaches by taking into consideration the interactions of entities with DeFi protocols and proposing novel features that improve the classification results.

\textbf{Contribution:} Overall our paper makes the following contributions:
\begin{enumerate}
    \item \textbf{Datasets:} we collected DeFi transactions across 23 protocols and 12 chains and created three datasets: i) a very comprehensive dataset thanks to a collaboration with Covalent \cite{noauthor_covalent_nodate} of all the DeFi-related transactions from the major DeFi protocols including Aave, Compound, Curve, Lido, and Yearn (but also including 18 further protocols); ii) compiled all the publicly available labeled datasets associated with illegitimate activities such as phish, hack, heist, and scam from the literature and through online resources. This set of malicious addresses consists of more than 10,000 address records; iii) identified all the accounts who participated in illegitimate activities and interacted with the major DeFi protocols. As a result, we are the first, to the best of our knowledge, to utilize a labeled dataset for Ethereum DeFi malicious accounts detection.
    \item\textbf{Features Extraction:} a set of features was extracted based on the Ethereum transactions and behavior of the entities. Additionally, we introduce a set of novel DeFi-related features, the extraction and exploitation of which improved the evaluation results.
     \item\textbf{Machine Learning Framework:} we train different ML methods to detect the malicious entities and evaluate their performance using the widely used metrics of Accuracy, Precision, Recall, F1-score and F2-score, while we tackle the data imbalance issue, which is quite common in this field, by employing oversampling techniques.
\end{enumerate}
The rest of this paper is structured as follows: In Section 2, we present the related work that has been done in the field, which formed the basis of our proposed solution. Section 3 contains the methodology followed, including the data collection process and the feature extraction, while in Section 4 we present the experimental results of the machine learning algorithms. Finally, Section 5 concludes with brief remarks on potential avenues for future research.

\section{Related Work}
\label{sec:Related-Work}
\noindent In the recent years, several research works leveraged machine learning methods to address the problem of Fraud Detection in the Bitcoin and Ethereum Blockchain platforms \cite{li2022survey}. 
Poursafaei et al. \cite{poursafaei2020detecting} extracted a set of features from the Ethereum blockchain data to represent the transactional behavior of entities. In their solution, they applied different Machine Learning classification algorithms (such as Logistic Regression, Support Vector Machine, Random Forest) in order to identify malicious entities in the Ethereum blockchain network.
Farrugia et al. \cite{farrugia2020detection} proposed a ML classification method based on XGBoost to detect illicit accounts and published their respective Ethereum addresses dataset including accounts that were flagged by the Ethereum community for illicit behavior for a number of cases, some of which include: (i) trying to imitate other contract addresses providing tokens, (ii) scam lotteries, (iii) fake initial coin offerings (ICO), (iv) imitating other users, (v) Ponzi schemes, (vi) phishing and (vii) mirroring websites. 
In a similar direction, Weber et al. \cite{weber2019anti} utilized Graph Convolutional Networks over a labelled dataset of ilicit Bitcoin transactions.

Other research works focused on Decentralized applications (DApps) and DeFi.
An analysis of the behavioral characteristics of the users of DApps was performed by \cite{min2022portrait} applying unsupervised Self-Organizing Map (SOM) based classification of addresses to distinguish blockchain investors and players. 
Darlin et al. \cite{darlin2022debt} proposed an Ethereum address grouping algorithm and a classification algorithm based on DeFi protocols to calculate the percentage of fund flows into DeFi lending platforms that can be attributed to debt created elsewhere in the system, analyzing data from five major DeFi protocols. 
Recent studies provided useful insights after analyzing Bitcoin/Ethereum and DeFi functions and data combined with external sources and datasets, such as the public sentiment from Twitter \cite{bouraga2022fork}, or  high-frequency volatility and price data  \cite{kyriazis2023monetary}.

Li et al. \cite{li2021measuring} used a set of 3559 labeled Ethereum addresses that have been marked as malicious, analyzed transactions at blocks from July 2015 until December 2020  and applied a network clustering analysis to identify whether additional Ethereum addresses should be marked as malicious. Trozze et al. \cite{trozze2023degens} used open-source investigative tools to study the DeFi frauds and money laundering, focusing on tokens, smart contracts, rug pulls and how tools can be used to extract evidence of scams on Ethereum. Wang et al. \cite{wang2022defiscanner} proposed a system called DeFiScanner, trained on a set of 50910 DeFi transactions, extracting 37 features about functions and transactions. They adapted word2vec and then used seven models to detect attacks, namely, SVM, k-means, autoencoder, deep autoencoder, LSTM, CNN, and LSTM-CNN. DeFiScanner only deals with DeFi attacks on the transaction layer and focuses on multilayer attacks on DeFi protocols. Their proposed extracted features are for each transaction such as the number of transfer functions, withdraw functions, and the related tokens, since one transaction might contain multiple events.

\begin{figure*}[ht]
\includegraphics[width=\linewidth]{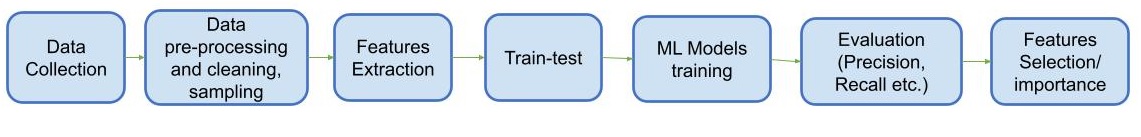}
\caption{Overview of the proposed solution - Machine Learning Framework pipeline } \label{figure-pipeline}
\end{figure*}

    In our study, we differentiate from the previous approaches since: i) we gathered a large dataset with more than 54,000,000 DeFi transactions of unique addresses interacting with major DeFi protocols (Aave, Compound, Curve, Lido, and Yearn) from May 2019 until March 2023; ii) use a large labeled set of addresses compiling all available public sources; iii) we extract features about addresses and more specifically 414 DeFi-related features and 9 transactional features in the classification process; iv) we focus in identifying a target set of 80 addresses labeled as DeFi malicious; v) we propose a Machine Learning Framework using SVM, Random Forest, Logistic Regression, XGBoost, and an Artificial Neural Network as shown in Fig.\ref{figure-pipeline}. Each of the components of this framework, is presented in detail in the following section.

\section{Methodology}
\label{sec:Methodology}

\begin{table}[h]
\centering
\begin{tabular}{>{\raggedright\arraybackslash}p{2cm}>{\raggedright\arraybackslash}p{5cm}}
\hline
Author/Source &  Description and Reference \\ \hline \hline
CryptoScamDB & Open-source dataset tracks malicious URLs and their associated addresses 
https://cryptoscamdb.org/
\\ 
Etherscan labels & A list of tagging, and categorizing of addresses and tokens listed on Etherscan \url{https://etherscan.io/}
\\ 
Hall et al.  & A labeled dataset with 5212 addresses \cite{hall_efficient_2021} \\ 
Tether Blacklisted Addresses & A list of Ethereum addresses \url{https://cointelegraph.com/news/tether-blacklists-39-ethereum-addresses-worth-over-46-million} \\ 
MyEtherWallet blacklist & A list of Ethereum addresses \url{https://github.com/MyEtherWallet/ethereum-lists/blob/master/src/addresses/addresses-darklist.json} \\ 
Ferrugia et al.  & A public dataset of 2179 Ethereum accounts \cite{farrugia2020detection}
\\ 
Xblockpro Ethereum Datasets & Various labeled datasets \url{http://xblock.pro/tx/} \\ 
Al-E’mari et al \cite{al-emari_labeled_2021} &  \url{https://github.com/salam-ammari/Labeled-Transactions-based-Dataset-of-Ethereum-Network} \\ \hline
\end{tabular}
\caption{Summary of datasets and sources available with annotated data of malicious addresses}
\label{table-datasets}
\end{table}

\subsection{Data Collection}
Data used for this study originate from Ethereum as well as publicly available datasets published by existing research studies and complementary publicly available web resources. 
Python and NodeJS, scripts were used for data retrieval and analysis and learning models are applied using the scikit-learn package\footnote{\url{https://scikit-learn.org/}}.
For initial experiments, we filtered for the relevant data from the Google BigQuery’s public Ethereum dataset. 
As described, the main data on the DeFi events were provided by Covalent\footnote{\url{https://www.covalenthq.com/}}.
We have also set up an Ethereum full archive node (8-core Intel i7-11700 CPU, 4.8GHz, 32GB RAM, 10TB SSD), in order to have a dedicated server for the purposes of this research, and to have a resource for running all the conducted experiments and extracting results. 
By utilizing the node, we have unlimited access to all the transactions that took place since the first Ethereum block in July 2015, also called the Genesis block and we were able to search for transactions including specific actions that characterize the applications and protocols of Decentralized Finance. 

We collected transactions occurring between blocks 5771740 (May 5, 2019 1:10:38 PM UTC) and 36843000 (March 12, 2023 4:46:11 PM UTC), inclusive, for 23 protocols including Aave, Compound, Curve, Lido, and Yearn. Each protocol has upgraded the initial versions of their smart contracts, and in some cases multiple versions of the protocol may exist at one time. Therefore, we collected data from the set of smart contracts associated to the different protocol versions.

\subsubsection{DeFi transactions dataset:}
We identified the smart contracts belonging to DeFi protocols. Following the data collection process described above, in this querying process for DeFi activities, we focused on addresses interacting with these DeFi protocols and the respective transactions. To this direction, we gathered all the transaction records through these contracts assembling them into the "DeFi transactions dataset", gathering more than 54,000,000 transactions for 550,000 unique entities (addresses). 


\subsubsection{Covalent dataset on decoded DeFi events:} Covalent (\url{https://www.covalenthq.com/}) is a new blockchain company that offers one of the most complete dataset on DeFi protocols through APIs to their clients. They have indexed more than 100B transactions across 60 blockchains and 200k smart contracts. On top, they clean and normalize the data, to make it more accessible for users. Covalent has graciously offered us to use their dataset for the research presented in this paper. The underlying dataset presented is hence one of the most advanced in the industry.

\subsubsection{Labeled Dataset}
\begin{figure*}[h]
\includegraphics[width=\linewidth]{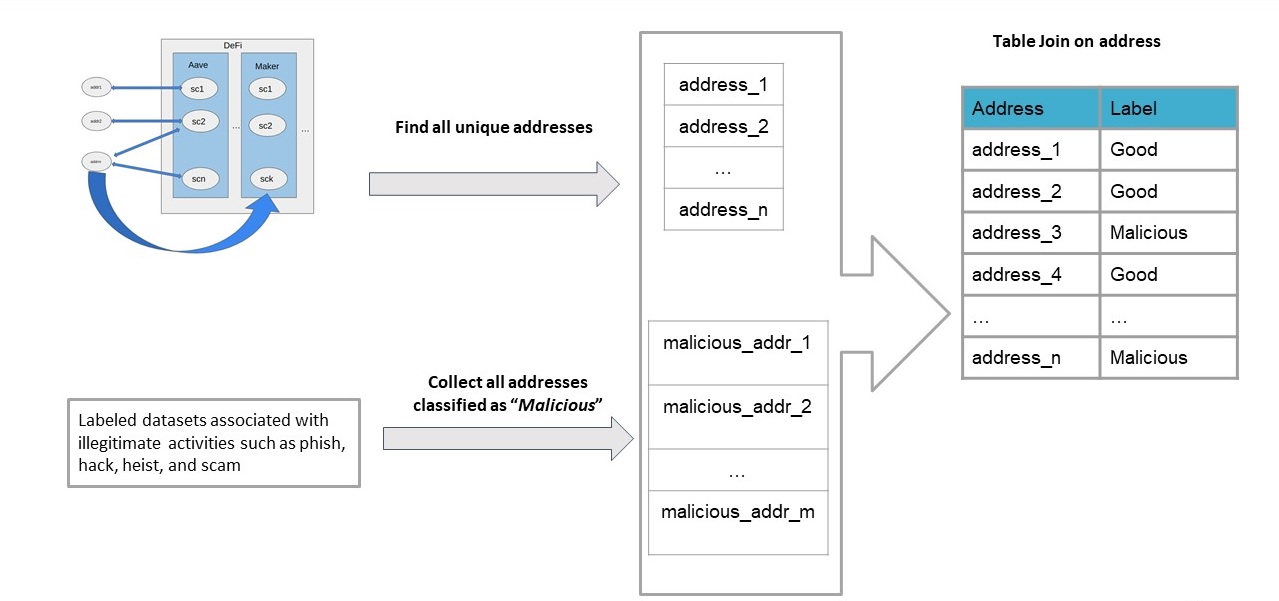}
\caption{Process of creating the labeled dataset for Ethereum DeFi fraud detection. An example of using CryptoScamDB is depicted. } 
\label{figure-annotation}
\end{figure*}

We compiled all the publicly available labeled datasets associated with illegitimate activities such as phish, hack, heist, and scam from the literature and through publicly accessed online resources.
In the Table \ref{table-datasets}, the online sources and available datasets are summarized and it consists of more than 10,000 addresses. Followingly, we identified all the accounts from the previous labeled dataset, which interacted with the major DeFi protocols from the "DeFi Transactions Dataset", creating a labeled dataset with addresses. While we retrieved more than 10,000 malicious addresses from the literature and previous works, only 81 addresses of them have also interacted with DeFi protocols. Figure \ref{figure-annotation} describes part of this process.

\subsection{Data Pre-processing}

We are tackling a classification problem with a dataset, which is imbalanced. From the set of good addresses, we select 10K addresses, while the malicious class consists of 81 addresses.
Several research works have processed imbalanced datasets with blockchain data \cite{alarab2022effect}, applying techniques such as SMOTE \cite{poursafaei2020detecting}. To this direction, we apply SMOTE, increasing the size of the smaller class in the training part of the pipeline.
We also applied Data Cleaning in certain cases, where we handled missing numerical values with the respective median.

\begin{table}[h]
\centering
\begin{tabular}{>{\raggedright\arraybackslash}p{.7cm}>{\raggedright\arraybackslash}p{6.5cm}}
\hline
Fea\-tures & Details \\ \hline \hline
1 & \textbf{1 calculation on 1 metric:} Total number of transactions \\
1 & \textbf{1 calculation on 1 metric:} Share of total submitted transactions to mempool successful  \\
2 & \textbf{2 calculations on 1 metric:} Standard deviation, and Max - Min (i.e., age) on block height (i.e., age)  \\
1 & \textbf{1 calculation on 1 metric:} Number of transactions (1st row) divided by age of wallet (3rd row)  \\
3 & \textbf{3 calculations on 1 metric:} Mean, max, and std on gas costs per transaction \\
\hline
8 &  Total number of transactional features \\ \hline
\end{tabular}
\caption{Extracted features from the transactional behavior of accounts}
\label{table-existing-features}
\end{table}

\subsection{Feature Extraction}
\begin{figure*}[h]
\centering
\includegraphics[width=1\textwidth]{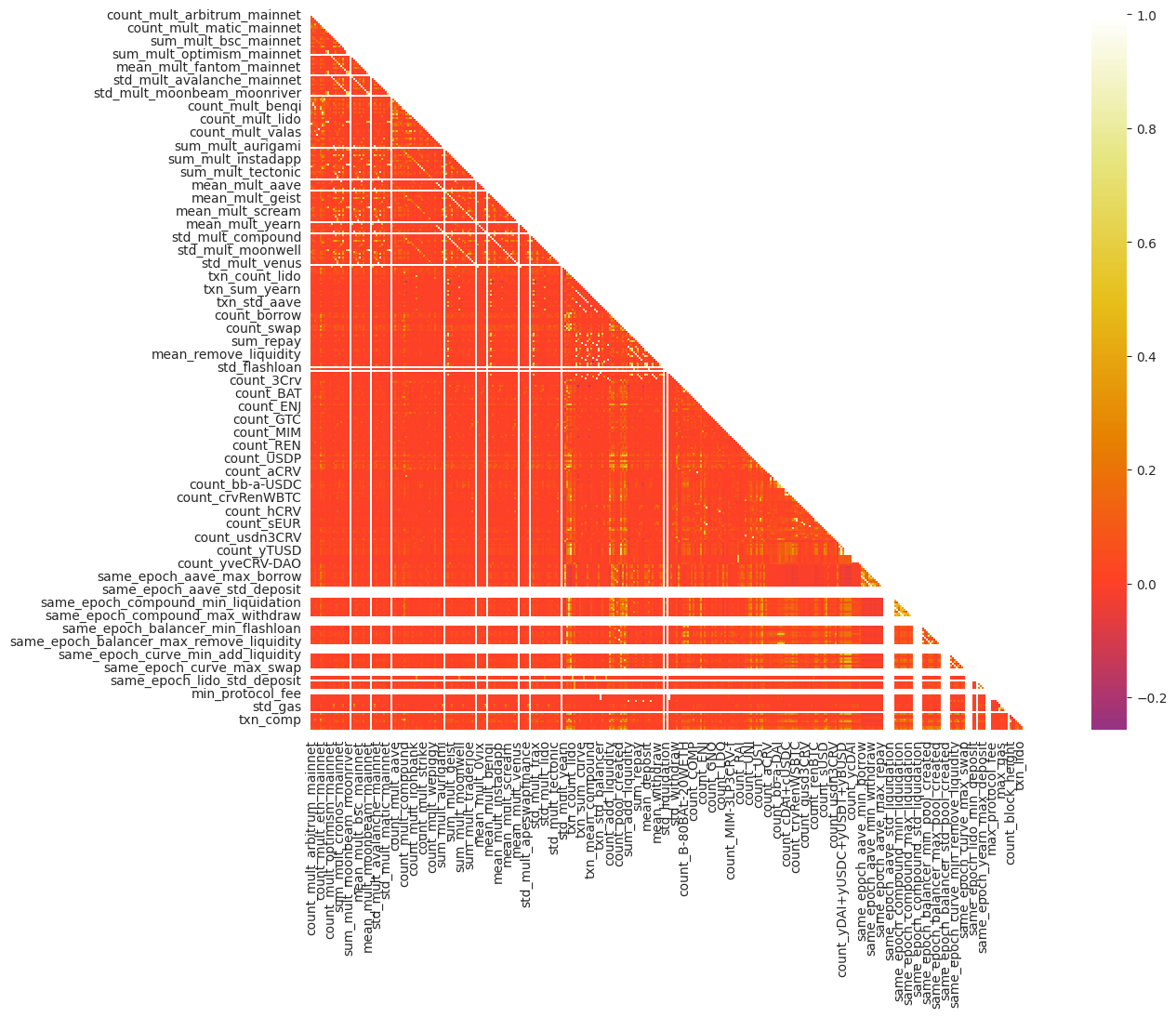}
\caption{Correlation Matrix including DeFi-related features} \label{figure-correlation-matrix-after}
\end{figure*}

In Table \ref{table-existing-features}, the list of extracted features based on the transactional behavior of entities is presented. Table \ref{table-defi-features} explains the extracted features based on the interactions of entities with DeFi protocols.
We also made sure that the features are not too correlated; the correlation can be found in the Figure \ref{figure-correlation-matrix-after}.

\begin{figure*}[h]
\centering
\includegraphics[width=1\textwidth]{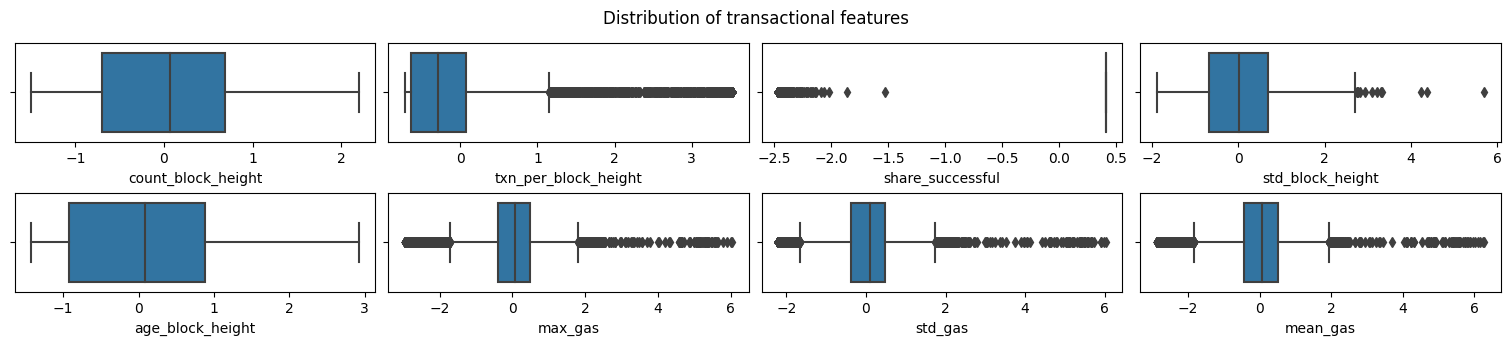}
\caption{Distribution of transactional features} \label{figure-distribution-features-before}
\end{figure*}

Additionally, we experimented with the normalization of features and filtering out features with zero variance. An investigation of the distribution of the features using boxplots was also performed, as shown in the Figure \ref{figure-distribution-features-before}.

\begin{table*}[h]
\centering
\begin{tabular}{>{\raggedright\arraybackslash}p{1.2cm}>{\raggedright\arraybackslash}p{14.5cm}}
\hline
Fea\-tures & Details \\ \hline \hline
24 & \textbf{3 calculations on 8 events:} Sum, mean, and std on 8 events: Add liquidity, remove liquidity, borrow, deposit, liquidation, repay, swap, withdraw \\ 
10 & \textbf{5 calculations on 2 types of fees:} Min, max, std, mean, and median on 2 types of fees: Protocol fees, and gas fees related to the DeFi protocol interaction \\
92 & \textbf{4 calculations on 23 protocols:} Number of transactions, sum, mean, and standard deviation of outgoing token value in USD on 23 protocols including Aaave, Compound, Yearn, Curve, Lido, Balancer (others are 0vix, Apeswapfinance, Aurigami, Bastion, Benqi, Frax, Geist, Granary, Instadapp, Ironbank, Moonwell, Radiant, Scream, Strike, Tectonic, Traderjoe, Valas, Venus, Wepiggy) \\
44 & \textbf{4 calculations on 11 chains:} Number of transactions, sum, mean, and standard deviation of outgoing token value in USD on 11 chains including Arbitrum, Aurora, Avalanche, BSC, Cronos, Ethereum, Fantom, Matic, Moonbeam, Moonbeam Moonriver, and Optimism \\
144 & \textbf{3 calculations on 8 events and 6 protocols:} Aggregate all transactions per every 1000 blocks (roughly 3.5h) to see activity in a timeframe and take min, max, and std on 8 events (add liquidity, remove liquidity, borrow, deposit, liquidation, repay, swap, withdraw) and 6 protocols (Aave, Balancer, Compound, Curve, Lido, and Yearn) \\
100 & \textbf{Top 100 tokens:} Number of transactions involving one of the top 99 most traded tokens in the sample, plus an aggregation of the long tail \\
\hline
414 &  Total number of DeFi features \\ \hline
\end{tabular}
\caption{Introducing DeFi-related features}
\label{table-defi-features}
\end{table*}

\subsection{Machine Learning Algorithms}

\begin{figure*}
\centering
\begin{subfigure}[b]{0.49\textwidth}
\includegraphics[width=\textwidth]{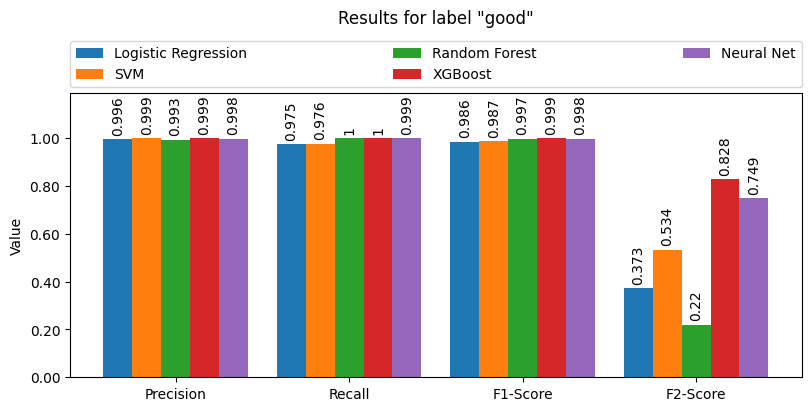}
\end{subfigure}
\begin{subfigure}[b]{0.49\textwidth}
\includegraphics[width=\textwidth]{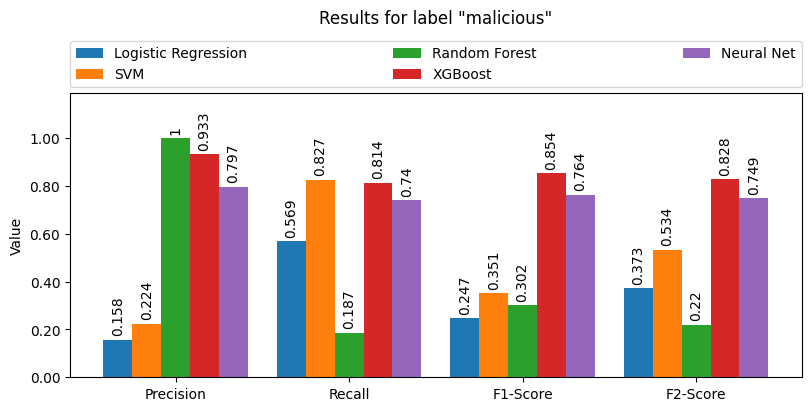}
\end{subfigure}
\caption{Evaluation results after utilizing DeFi-related features: minority class in the right part of the figure} \label{figure-Good-Bad-evaluation-after-features}
\end{figure*}

We used five different machine learning algorithms: Logistic Regression, Random Forest, Support Vector Machine (SVM), XGBoost, and an Artificial Neural Network (ANN).
For each of the models, we apply a 5-fold cross-validation (CV), hence splitting the dataset in 80\% training and 20\% testing data five times, and training the model on each batch independently. The presented results in Figure \ref{figure-Good-Bad-evaluation-after-features} is the average of this 5-fold CV. 

\begin{itemize}
\item \textbf{Logistic Regression, Random Forest and XGBoost:} Standard implementation with sklearn.
\item \textbf{SVM:} Standard implementation with sklearn and rbf kernel.
\item \textbf{ANN:} Implementation with TensorFlow\cite{tensorflow2015-whitepaper} with mean squared error (MSE) loss function and Adam optimizer. The model consists of five layers: a dense layer with 40 nodes and ReLU activation, a 0.3 dropout layer, a dense layer with 10 nodes and ReLU activation, a dense layer with 5 nodes and ReLU activation, and a dense output layer with 1 node and sigmoid activation.
The model is then trained through 30 epochs with 128 batch size. 
\end{itemize}

\section{Experimental Results}
\label{sec:experimental-results}

\begin{figure*}
\centering
\begin{subfigure}[b]{0.49\textwidth}
\includegraphics[width=\textwidth]{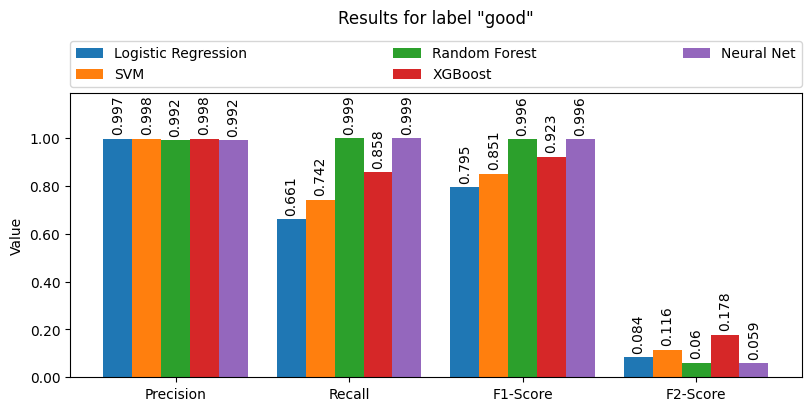}
\end{subfigure}
\begin{subfigure}[b]{0.49\textwidth}
\includegraphics[width=\textwidth]{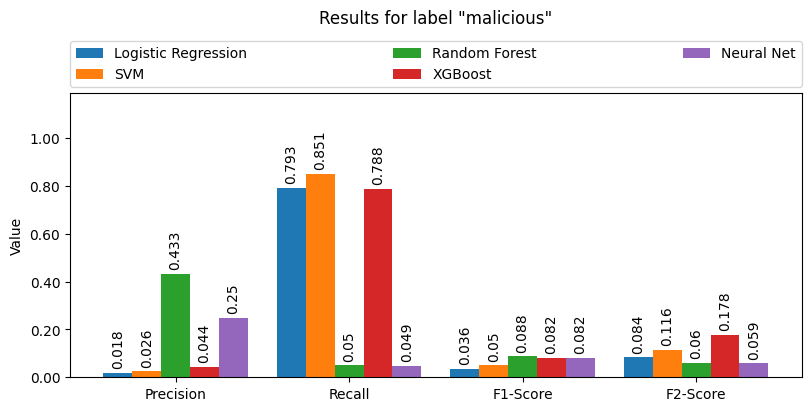}
\end{subfigure}
\caption{Evaluation results using only the transactional features for three classifiers: minority class in the right part of the figure} 
\label{figure-good-bad-evaluation-before-features}
\end{figure*}

\subsection{Model Verification Techniques}
To evaluate the performance of our models, we use the following widely used metrics: Precision, Recall, Accuracy, F1-score. They are defined through combinations of True Positive (TP), True Negative (TN), False Positive (FP), and False Negative (FN). Moreover, we decided to utilize the F2-score, in order to put more attention to the minority class, by minimizing False Negatives (erroneously classifying "Malicious" samples as "Good") over minimizing False Positives.

\begin{equation}
\textrm{Precision} = \frac{\textrm{TP}}{\textrm{TP} + \textrm{FP}}
\end{equation}

\begin{equation}
\textrm{Recall} = \frac{\textrm{TP}}{\textrm{TP} + \textrm{FN}}
\end{equation}

\begin{equation}
\textrm{Accuracy} = \frac{\textrm{TP} + \textrm{TN} }{\textrm{TP} + \textrm{TN} + \textrm{FP} + \textrm{FN} }
\end{equation}

\begin{equation}
\textrm{F1-score} = 2 \cdot \frac{\textrm{Precision} \cdot \textrm{Recall} }{ \textrm{Precision} + \textrm{Recall}}
\end{equation}

\begin{equation}
\textrm{F2-score} = \frac{5 \cdot \textrm{Precision} \cdot \textrm{Recall} }{ 4 \cdot \textrm{Precision} + \textrm{Recall}}
\end{equation}

\subsection{Evaluating the performance of the classification algorithms }
In this section, the effectiveness of the preprocessing and feature extraction is evaluated, while the results before and after including the DeFi-related features are compared. Figure \ref{figure-Good-Bad-evaluation-after-features} summarizes the performance of the classifiers.

It is apparent that all classifiers work well to classify the "Good" labels with high precision, recall, and F1-Scores. For the "Malicious" label, there are significant differences. Overall, XGBoost and the Neural Net have the highest F1- and F2-Scores. They hence perform best combining precision and recall. Random Forest has a high precision, however a low recall, whereas the SVM performs the other way. Logistic Regression performs the worst overall.

\subsection{Feature importance}
\begin{figure*}[t]
\centering
\includegraphics[width=1\textwidth]{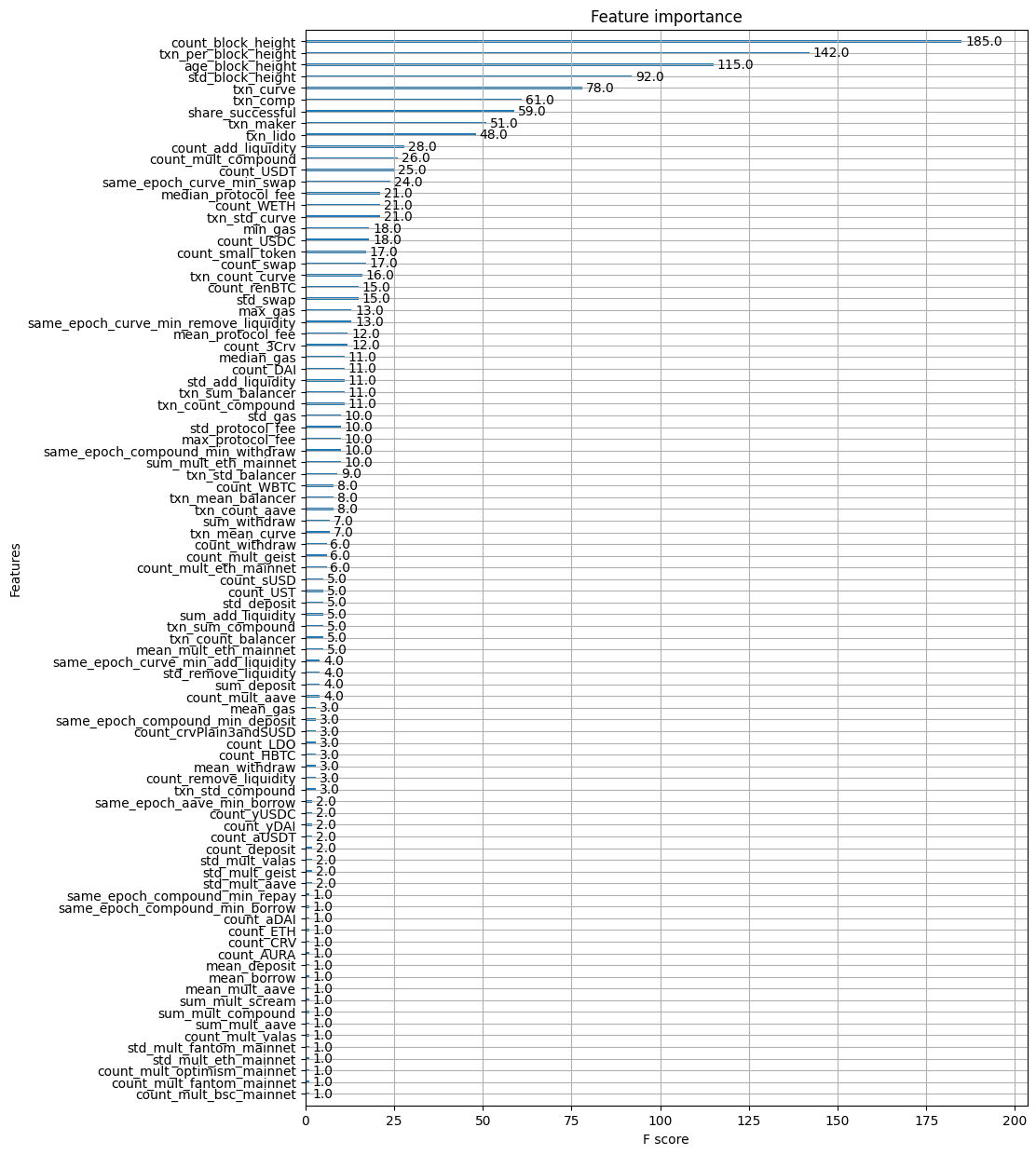}
\caption{Features importance after utilizing DeFi-related features} \label{features-importance-after}
\end{figure*}

We visualized the importance of each feature of the best model. It was inferred that the most important feature for this task is in DeFi-related, while other DeFi-rlated features have bigger impact than traditional transactional features. Figure \ref{features-importance-after} shows the features importance after utilizing the new features for the XGBoost classifier, confirming the significance of the DeFi features. This is also shown in Figure \ref{figure-good-bad-evaluation-before-features}, where we ran the same models on only the transactional features.

\subsubsection{Transactional features:} Overall, the number of transactions, the age of the wallet, and transactions over time are the most relevant features. 

\subsubsection{DeFi features:} As discussed above, the DeFi features have a strong impact on the results and classification. No feature stands out in particular, and instead many of the engineered features are used in the classifier. This suggests that the classifier is utilizing information that is not immediately apparent, unlike e.g., the transactional features. This can help in the future to detect malicious wallets before they are tagged as such.

\section{Conclusion}
\label{sec:conclusion}

In this paper, we studied how Machine Learning classification methods can be used to detect entities that interacted with DeFi protocols and also have been labeled in dedicated Ethereum annotated datasets. We gathered large datasets of transactions focusing on the major DeFi protocols.
We observed that the introduction of DeFi-related features significantly improved the performance of the algorithms, especially for the minority class of the dataset.
The authors would like to note that they only analyze publicly available data from existing works in this field as well as public online sources (websites) and do not take part in any labeling or identification processes.

It was found that XGBoost and a Neural Network are the best classifiers for malicious DeFi addresses, with a Precision of 0.80-0.93, a Recall of 0.74-0.85, and an F1-Score of 0.76-0.85 (for NN vs. XGB respectively). It was also shown that the 400+ DeFi related features improve the classification accuracy significantly; the same models for only transactional non-DeFi related features yields an F1-Score of 0.08 respectively.

Regarding future extensions of our system, we consider exploring more machine learning algorithms and more specifically Deep Learning algorithms. Additinal preprocessing techniques will be also integrated within the Machine Learning Framework pipeline and additional methods for oversampling and undersampling will be also considered.

\clearpage 
\onecolumn
\bibliography{references}

\end{document}